\documentclass[11pt]{emulateapj}
\usepackage{graphicx}
\usepackage{multirow}
\usepackage{textcomp}
\usepackage{epsfig}
\usepackage{amsmath}
\usepackage{amssymb}
\usepackage{amsthm}
\usepackage{xy}

\def\src{SGR\,0418$+$5729}
\def\xin{RX\,J0720.4$-$3125}

\newcommand{\xmm}{{\em XMM--Newton}}

\def\nh {$N_{\rm H}$}
\def\chisq {$\chi ^{2}$}
\def\rchisq {$\chi_{\nu} ^{2}$}
\def\ergs {erg\,s$^{-1}$}
\def\ergscm2 {erg\,s$^{-1}$cm$^{-2}$}

\def\cm2 {cm$^{-2}$}
\def\arcsec{$^{\prime\prime}$}

\shortauthors{A. Borghese et al.}

\begin{document}

\title{Discovery of a strongly phase-variable spectral feature in the isolated neutron star RX J0720.4$-$3125}

\author{A. Borghese\altaffilmark{1}, N. Rea\altaffilmark{1,2}, F. Coti Zelati\altaffilmark{1,3,4}, A. Tiengo\altaffilmark{5,6,7}, R. Turolla\altaffilmark{8,9}}

\altaffiltext{1}{Anton Pannekoek Institute for Astronomy, University of Amsterdam, Postbus 94249, NL--1090 GE Amsterdam, the Netherlands.}
\altaffiltext{2}{Institute of Space Sciences (CSIC--IEEC), Carrer de Can Magrans S/N, 08193 Barcelona, Spain.}
\altaffiltext{3}{Universit\`a dell'Insubria, via Valleggio 11, I--22100 Como, Italy}
\altaffiltext{4}{INAF--Osservatorio Astronomico di Brera, via Bianchi 46, I--23807 Merate (LC), Italy}
\altaffiltext{5}{INAF--Istituto di Astrofisica Spaziale e Fisica Cosmica, via E. Bassini 15, I--20133 Milano, Italy}
\altaffiltext{6}{Istituto Universitario di Studi Superiori, piazza della Vittoria 15, I--27100 Pavia, Italy}
\altaffiltext{7}{Istituto Nazionale di Fisica Nucleare, Sezione di Pavia, via A. Bassi 6, I--27100 Pavia, Italy}
\altaffiltext{8}{Dipartimento di Fisica e Astronomia, Universit\`a di Padova, via F. Marzolo 8, I--35131 Padova, Italy}
\altaffiltext{9}{Mullard Space Science Laboratory, University College London, Holmbury St. Mary, Dorking, Surrey RH5 6NT, UK}

\begin{abstract}
We present the discovery of a strongly phase-variable absorption
feature in the X-ray spectrum of the nearby, thermally-emitting,
isolated neutron star \xin. The absorption line was detected
performing detailed phase-resolved spectroscopy in 20 {\xmm}
observations, covering the period May 2000 -- September 2012. The
feature has an energy of  $\sim$750\,eV, an equivalent width of
$\sim30$\,eV, and it is significantly detected  for only $\sim
20$\% of the pulsar rotation. The absorption feature appears to be
stable over the timespan covered by the observations. Given its
strong dependence on the pulsar rotational phase and its narrow
width, a plausible interpretation is in terms of resonant proton
cyclotron absorption/scattering in a confined magnetic structure
very close to the neutron star surface. The inferred field in such
a magnetic loop is $B_{\rm loop}\sim 2\times10^{14}$\,G, a
factor of $\sim7$ higher than the surface dipolar magnetic field.
\end{abstract}


 \section{Introduction}

\begin{table*}
\begin{center}
\caption{Summary of the {\xmm}/EPIC-pn observations of RX J0720.4$-$3125$^a$}
\label{tab:log}
\footnotesize{
\begin{tabular}{@{}lclccc}
\hline
\hline
Obs. ID         &  Obs. Date        & Read-out mode / filter    & Live time     & Source net count rate     & Pile-up fraction ratios           \\
            & YYYY-MM-DD    &                   & (ks)          & (counts s$^{-1}$)             & $r$=0"--30''  \\
\hline
0124100101  & 2000 May 13   & FF / thin             & 42.8      & 6.46(1)               & 0.963(3) \\
0132520301  & 2000 Nov 21       & FF / medium           & 22.7      & 5.60(2)               & 0.964(4)      \\
0156960201  & 2002 Nov 06       & FF / thin             & 25.6      & 6.60(2)               & 0.969(3)      \\
0156960401  & 2002 Nov 08       & FF / thin             & 27.1      & 6.54(2)               & 0.966(3)  \\
0158360201  & 2003 May 02   & SW / thick            & 51.0      & 3.480(8)          & 1.011(3)  \\
0161960201  & 2003 Oct 27       & SW / thin         & 12.6      & 7.52(2)               & 1.013(5)      \\
0164560501  & 2004 May 22   & FF / thin             & 32.0      & 6.96(1)               & 0.971(3)      \\
0300520201  & 2005 Apr 28       & FF / thin             & 38.1      & 6.86(1)               & 0.968(3)   \\
0300520301  & 2005 Sep 22       & FF / thin             & 39.1      & 6.93(1)               & 0.969(3)      \\
0311590101  & 2005 Nov 12       & FF / thin             & 33.5      & 6.75(1)               & 0.970(3)      \\
0400140301  & 2006 May 22   & FF / thin             & 17.6      & 6.83(2)               & 0.970(4)      \\
0400140401  & 2006 Nov 05       & FF / thin             & 17.6      & 6.90(2)               & 0.966(4)      \\
0502710201  & 2007 May 05   & FF / thin             & 17.4      & 6.80(2)               & 0.968(4)      \\
0502710301  & 2007 Nov 17       & FF / thin             & 20.1      & 7.71(2)               & 0.971(4)      \\
0554510101  & 2009 Mar 21       & FF / thin             & 16.7      & 6.84(2)               & 0.967(4)  \\
0601170301  & 2009 Sep 22       & FF / thin             & 15.0      & 6.77(2)               & 0.968(4)      \\
0650920101  & 2011 Apr 11       & FF / thin             & 17.6      & 6.61(2)               & 0.973(4)      \\
0670700201  & 2011 May 02   & FF / thin             & 23.6      & 6.73(2)               & 0.965(3)      \\
0670700301  & 2011 Oct 01       & FF / thin             & 22.2      & 6.60(2)               & 0.972(3)      \\
0690070201  & 2012 Sep 18       & FF / thin             & 22.3      & 6.60(2)               & 0.970(3)      \\
\hline
\hline
\end{tabular}}
\begin{list}{}{}
\item[$^{a}$] FF: full-frame (time resolution of 73 ms); SW: small
window (time resolution of 6 ms). Live time refers to the duration
of the observations after filtering for background flares (see
text). Count rates refer to the spectra extracted within a circular region with $\textsc{pattern} =
0$. Errors on the
count rates are quoted at the $1\sigma$ confidence level. Pile-up
fraction ratios were calculated for single events alone and in the
0.1--1.2 keV energy range using the {\tt SAS epatplot} tool.
\end{list}
\end{center}
\end{table*}


\xin\ belongs to a group of seven thermally-emitting, radio-quiet,
and nearby ($\lesssim$ 500 pc) isolated neutron stars, originally
discovered in \textit{ROSAT} all-sky survey data, and often
referred to as the X-ray Dim Isolated Neutron Stars (XDINSs). They
are among the closest known neutron stars, and are characterized
by X-ray luminosities $L_{\rm X}\approx 10^{30-33}$\ergs, long
spin periods ($P\sim 3$--$11$\,s) and inferred surface dipolar
magnetic fields $B_{\rm dip}\approx 10^{13}$\,G, partially
overlapping those of the magnetars (see van Kerkwijk \& Kaplan
2007 and Turolla 2009 for reviews). XDINSs have estimated ages of
a few 10$^5$\,yr, derived from cooling curves and kinematics,
while the characteristic ages are somewhat longer
($\sim$10$^6$\,yr). The X-ray spectra show thermal emission with
inferred temperatures  $kT \approx$ 50--100\,eV and lack the non-thermal,
power-law component often observed in other isolated
neutron stars. Although a blackbody model provides an overall good
description of their X-ray spectra, broad absorption features have
been found in six of the XDINSs with \xmm\ observations. Their
origin is unclear: they can be produced either by proton cyclotron
resonances/atomic transitions in a magnetized atmosphere, or by an
inhomogeneous surface temperature distribution, as recently
suggested by Vigan\`o et al. (2014).

\xin, the second brightest member of the class, was discovered by
Haberl et al. (1997) as an isolated, pulsating neutron star with
an 8.39\,s spin period. The X-ray spectrum is best modelled by a
blackbody ($kT\sim85$\,eV) plus a broad ($\sim 70$\,eV wide) absorption
feature centered at $\sim$ 270\,eV (Haberl et al. 2004). A further, narrow ($\sim 5$\,eV wide) line was identified in \xmm\ Reflection Grating Spectrometer (RGS)
data at $\sim$ 570\,eV (Hohle et al. 2012a and references therein), possibly due to Oxygen circumstellar/interstellar absorption.
Unlike other
XDINSs, \xin\ exhibits long-term variations in its timing and
spectral parameters. In the period 2001--2003 the total flux
stayed almost constant, whereas the blackbody temperature
increased from $\sim$ 84 to over 94\,eV, and also the pulse
profile, the blackbody radius, and the equivalent width of the
absorption feature changed. In the following years this trend
appeared to reverse, with a decrease of the surface temperature.
Two main explanations for this behaviour were proposed: either
free precession, which predicts a cyclic pattern (Haberl et al.
2006), or a glitch that occurred around MJD = 52866$\pm$73\,days
(van Kerkwijk et al. 2007). According to the most recent analyses
(Hohle et al.  2009, 2012b), the interpretation in terms of a
single sudden event around the proposed glitch epoch seems
favoured.

In this Letter we reanalyze all the archival \xmm\ observations of
\xin, performing a detailed phase-resolved spectral analysis. In
Section \ref{analysis} we  report on the data analysis, and
present the discovery of a second, new,  phase-dependent
absorption feature in the X-ray spectrum, the possible origin of
which is discussed in Section \ref{discussion}.



\begin{figure}
\begin{center}
\includegraphics[width=9cm]{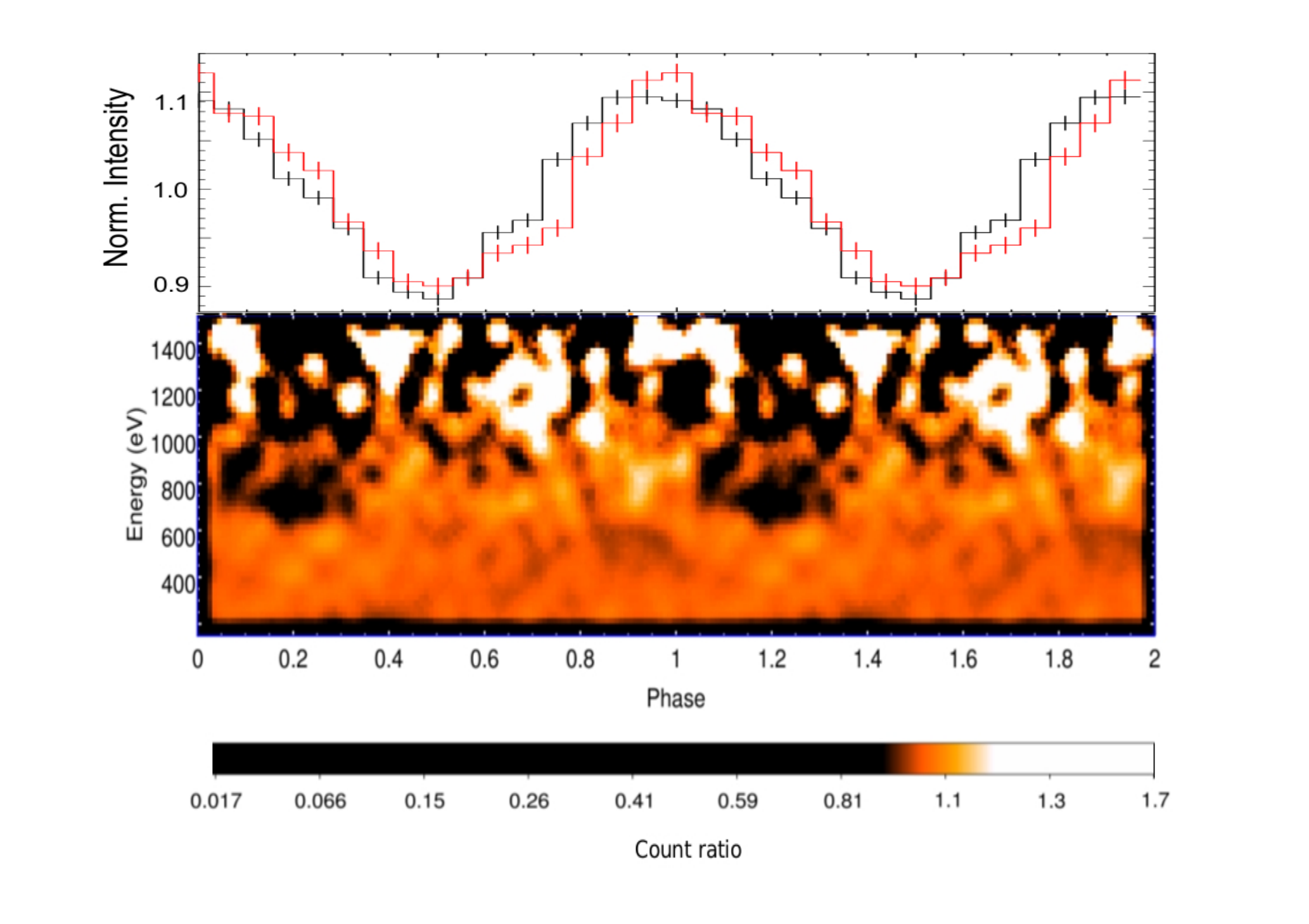}
\end{center}
\caption{{\em Top panel}: pulse profile for the May 2nd, 2003
observation (red) and for all observations merged together
(black). {\em Bottom panel}: normalized energy versus phase
image obtained by binning the EPIC source counts into 0.01 phase
bins and 25 eV energy channels for the observation performed on May 2nd, 2003. } \label{fig:fiamme}
\end{figure}

\begin{figure*}
\begin{center}
\includegraphics[width=15 cm]{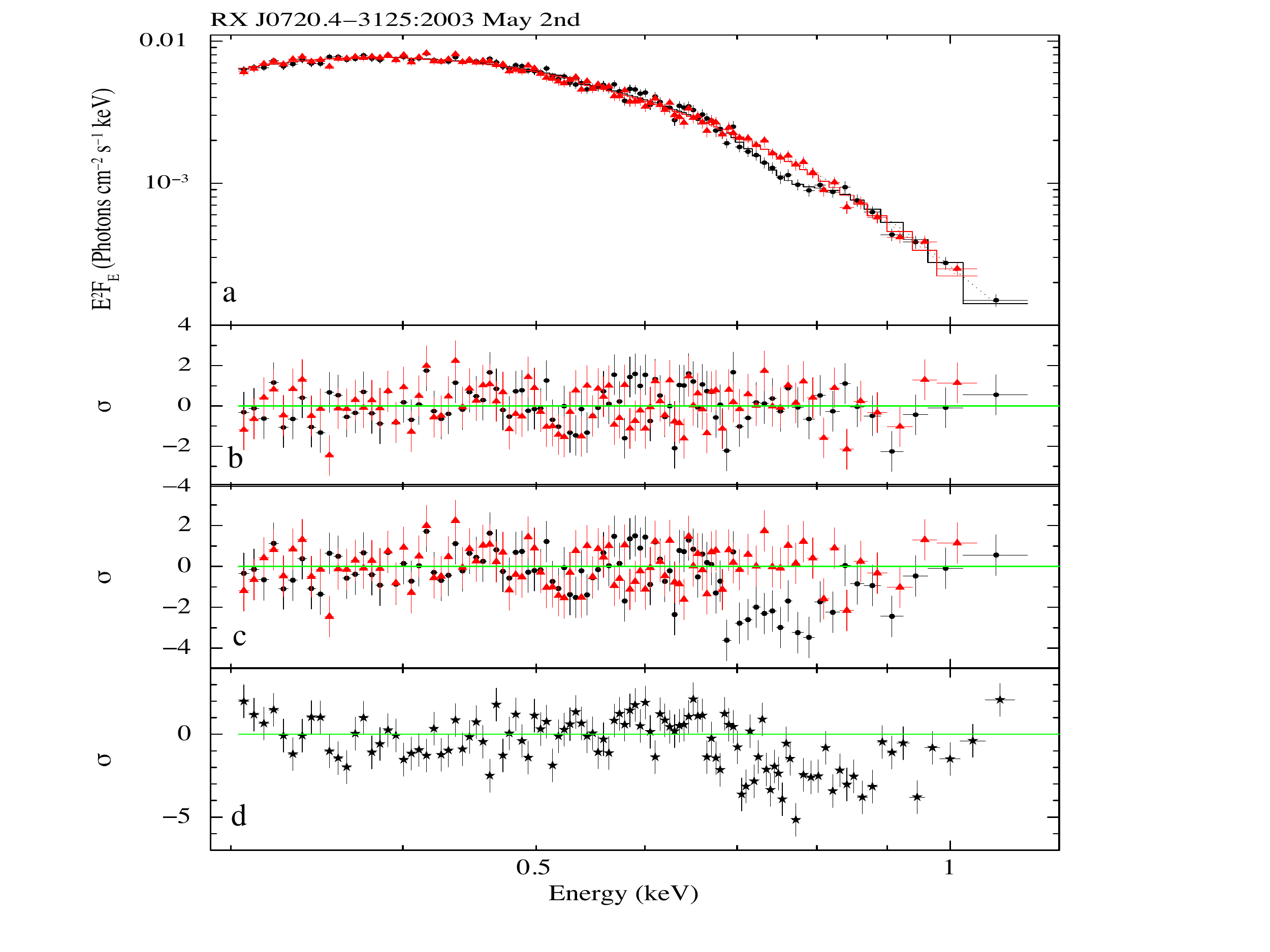}
\end{center}
\caption{From top to bottom. Panel \textit{a}: spectrum of the
0.1-0.3 phase interval (black circles) fitted with an absorbed
blackbody plus a Gaussian profile (\texttt{gauss} model); the
spectrum in the phase range 0.5-0.7 together with the best fitting
model (\texttt{tbabs*bbodyrad}) is also shown (red triangles).
Both spectra are from the May 2nd, 2003 observation. Panel
\textit{b}: residuals with respect to these models. Panel
\textit{c}: residuals of the previous spectra after setting the
line normalization to zero. Panel \textit{d}: residuals of the
spectrum relative to phase 0.1-0.3 after merging the phase-resolved spectra of all observations listed in Table\,1.} \label{fig:fit}
\vskip -0.1truecm
\end{figure*}
\begin{figure}
\includegraphics[width=6.4cm,height=5cm]{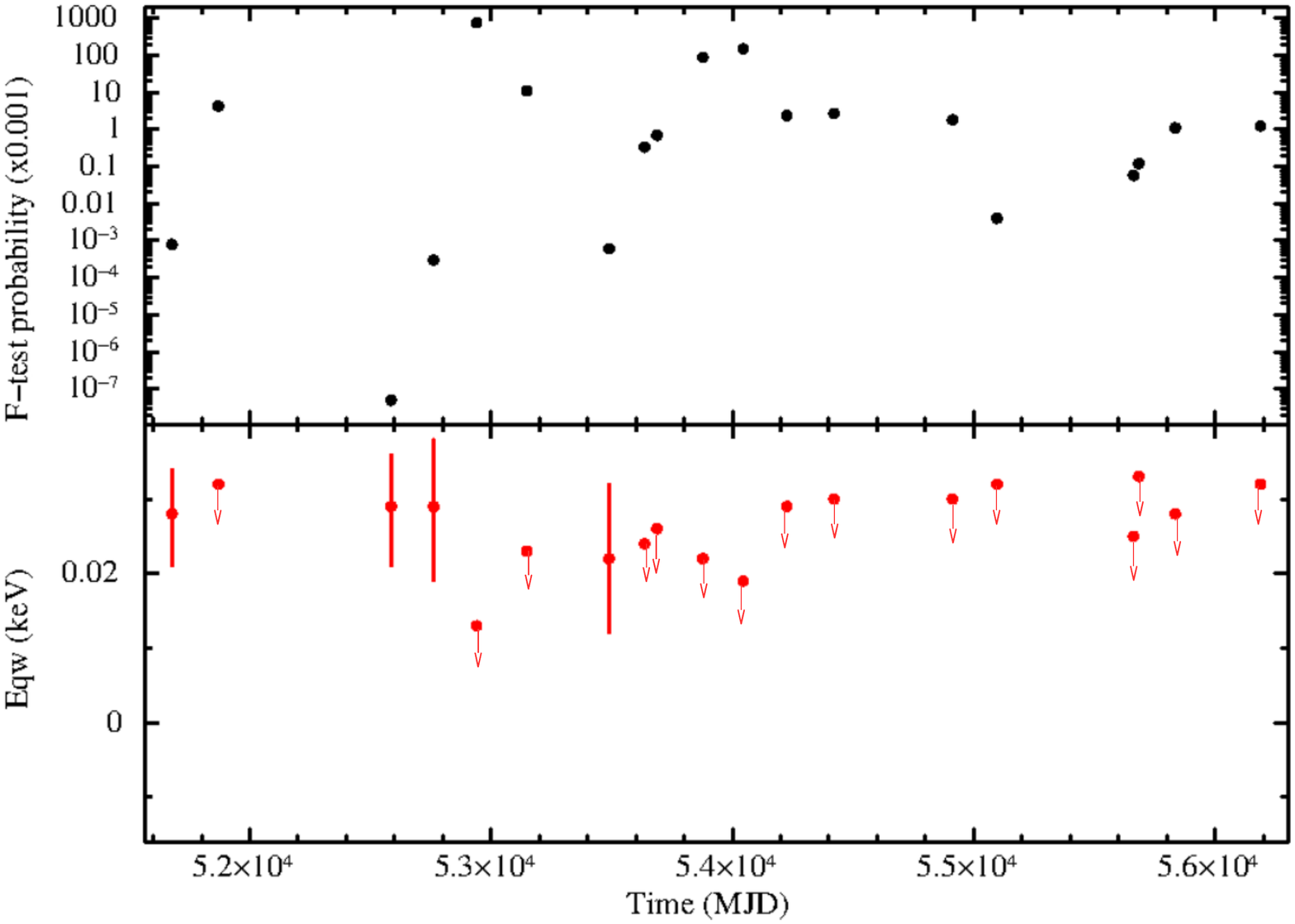}
\includegraphics[width=7cm,height=4.8cm]{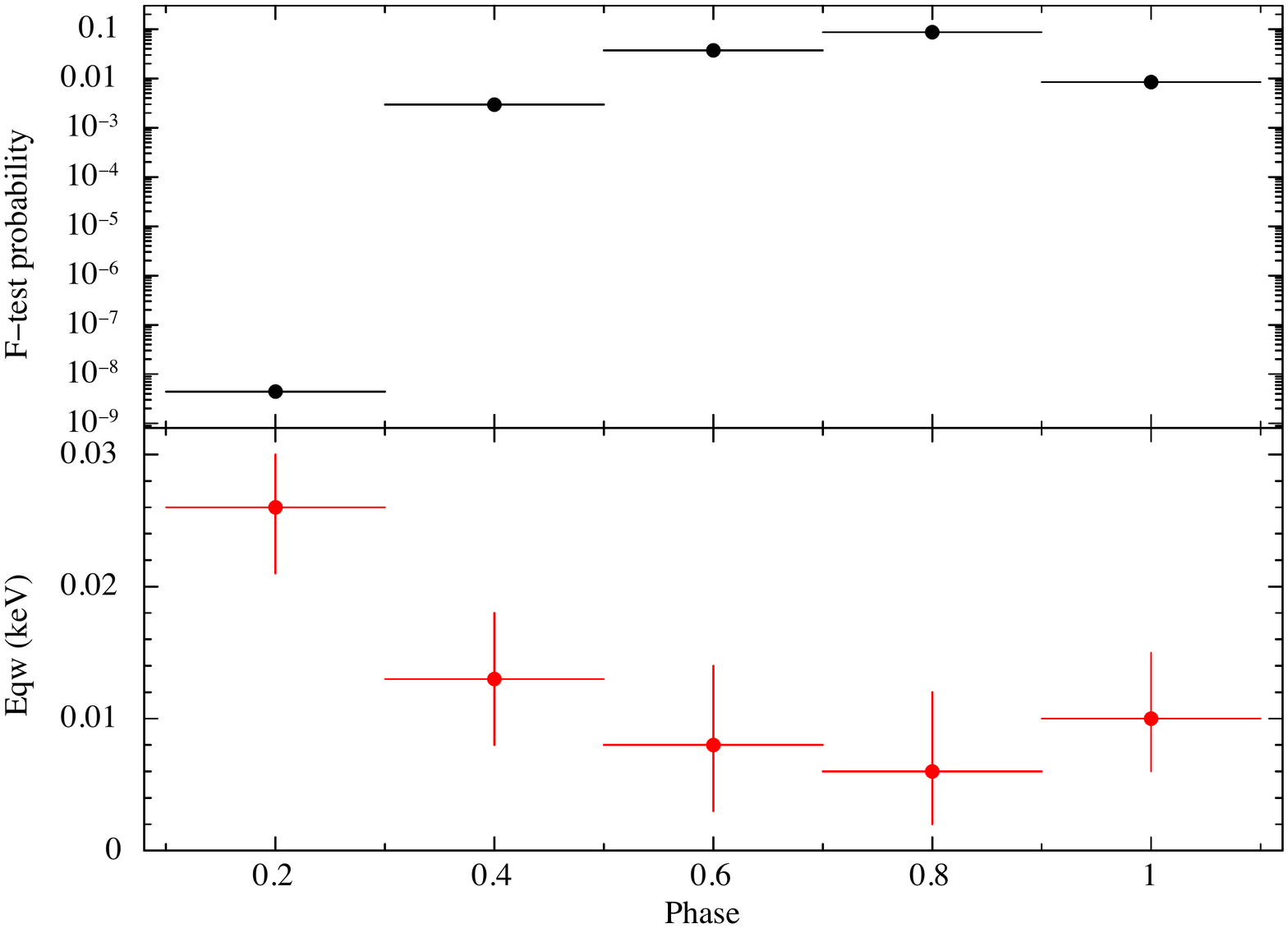}
\caption{\textit{Top panel}: evolution of the F-test probability
and line equivalent width with time for the spectra in the
0.1--0.3 phase range for all \xmm\ observations (the two
observations performed during November 2002 were merged being only
two days apart). \textit{Bottom panel}: F-test probability and
equivalent width as a function of the spin phase for the
spectrum of all merged observations.} \label{fig:ftest}
\end{figure}

\section{XMM-Newton data analysis and results}
\label{analysis}

\xin\ was observed 20 times by the {\xmm} satellite using the
European Photon Imaging Camera (EPIC). We used here only data
obtained with the EPIC-pn camera (Str\"{u}der et al. 2001) because
less affected by pile-up (see Table \ref{tab:log}). We processed
the data using the Science Analysis Software (\textsc{sas},
version 13.5.0), adopting the most recent calibration files
available, and we removed any particle flare via
good-time-intervals. All arrival times were referred to the Solar
System barycenter (source coordinates $\rm RA=07^h20^m24\fs961$,
$\rm Dec= -31^\circ25'50\farcs21$, J2000.0; Kaplan et al. 2003),
and a rotational phase was assigned to the source counts of all
the observations using the timing solution of Hohle et al. (2012b).
For the source distance we assumed $D = 286$\,pc (Tetzlaff
et al. 2011). All errors are reported at the 90\% confidence
level.

Spectral analysis was performed using the {\tt XSPEC} analysis
package (version 12.8.2, Arnaud 1996), using a minimum of 100
counts per bin for each spectrum, the \chisq\ statistics and a maximum oversampling of the spectral energy resolution of a factor of 3. The
fits were restricted to energies between 0.3 and 1.2 keV to exclude the broad absorption line at $\sim$270\,eV so as to reduce
the number of degrees of freedom (dof). We extracted the source spectra
from a circular region of radius 30$^{\prime\prime}$ centered on
the source point spread function (PSF) and the background counts from a circle of the same
size far from the source and on the same CCD. For Small
Window and Full Frame observations the pile-up level, estimated by
means of  the \textsc{sas} \texttt{epatplot} tool, was $<$1.5\%
and 4\%, respectively (see Table \ref{tab:log}). To mitigate the
pile-up we restricted our  spectral analysis to photons having
\textsc{flag} = 0 and $\textsc{pattern} = 0$.

We started our analysis from the May 2nd, 2003 observation, this being the longest observation, and least
affected by pile-up (the pn was operating in Small Window mode). The phase-averaged spectrum is well
fitted by an absorbed\footnote{We took into account the effects of
interstellar absorption along the line of sight through the
\texttt{TBABS} model with the \texttt{vern} cross sections (Verner
et al. 1996) and the \texttt{wilm} abundances (Wilms et al.
2000).} blackbody (reduced chi-square \rchisq\ = 1.17 for 130 dof)
yielding the following parameters: column density \nh\ $=
1.9(3)\times10^{20}$\cm2 , blackbody temperature $kT_{\rm BB}$
= 82.1$\pm$0.6\,eV and blackbody radius $R_{\rm BB}$ =
6.3$\pm$0.6\,km.

To study in detail the spectral variability with phase, we
produced a phase-energy image by binning the EPIC-pn source counts
into 100 rotational phase channels and 25-eV-wide energy channels,
and then normalizing to the phase-averaged energy spectrum. The
image (Fig. \ref{fig:fiamme}) shows a feature in the phase
interval 0.1--0.3, which gives a strong hint for the presence of
an absorption line at $\sim$750\,eV in the corresponding spectrum.
Above $\sim$ 1 keV the source counts become background dominated.

To better investigate the significance of the feature appearing in
Fig.\,1, we performed a phase-resolved spectral analysis for the May 2nd, 2003 observation dividing the rotational cycle in five
equal phase bins. For the spectrum relative to phase 0.1--0.3 the
addition of a Gaussian line in absorption to the blackbody
continuum leads to a significant improvement in the shape of residuals,
while a simple blackbody model gives an acceptable fit for all the
other phase bins. Both the \texttt{gabs} and \texttt{gauss} models
result in \rchisq\ = 0.88 for 89 dof,
corresponding to an F-test probability of 3.5$\times$10$^{-7}$. We
multiplied this value by 5, considering the five phase-intervals
as trials. In this way we obtained a significance of
$\sim5\sigma$. The values for line energy and width, obtained from
the two models, are compatible within the errors. The best-fit \texttt{tbabs*(bbodyrad +
gauss)} model gave the following parameters:  $kT_{\rm BB}$ = 83.0$\pm$0.1 eV, $R_{\rm BB}$ = 6.3$\pm$0.3 km, line energy E$_{line}$ = 745$^{+17}_{-27}$ eV, width $\sigma$ = 42$^{+51}_{-33}$ eV and normalization of -9.2$^{+3.5}_{-9.2}\times10^{-5}$ (the column density was frozen at the value obtained for the phase-averaged spectrum). The equivalent width of the feature is 28$^{+9}_{-11}$ eV.
Table 2
summarizes the results of the phase-resolved spectroscopy for the May 2nd, 2003 observation using the \texttt{tbabs*(bbodyrad +
gauss)} model. Fig. \ref{fig:fit} shows the phase-resolved
spectrum relative to phase 0.1--0.3 (black) in comparison with
that at phase 0.5--0.7 (red): the absorption feature clearly shows
up in the residuals.

\begin{table*}
\begin{center}
\caption{Pulse phase spectroscopy for May 2nd, 2003 observation. }
\begin{tabular}{lccccc}
\hline \vspace{0.2cm}
Parameter$^{a}$      & 0.1--0.3  &   0.3--0.5 & 0.5--0.7 & 0.7--0.9 & 0.9--1.1   \\
\hline
\multicolumn{6}{c}{BB} \\
\hline

   $kT_{\rm BB}$ (eV) & 81.1(6) & 81.9(6) & 82.2(6) & 82.6(6) & 82.5(9) \\

  $R_{\rm BB}$ (km) & 6.4$^{+0.6}_{-0.5}$ & 6.0$^{+0.6}_{-0.5}$ & 6.1$^{+0.6}_{-0.5}$ & 6.4$^{+0.6}_{-0.5}$ & 6.4$^{+0.7}_{-0.6}$  \\

Flux$^b$ & 0.86(1) & 0.82(1) & 0.86(1) & 0.95(1) & 0.95(1) \\

Unabs. Flux$^b$  & 1.07 & 1.01 & 1.06 & 1.17 & 1.17 \\

NHP $^d$ &  7.2$\times 10^{-2}$ & 5.9$\times 10^{-1}$ & 6.9$\times 10^{-1}$ & 4.2$\times 10^{-2}$ & 2.7$\times 10^{-1}$ \\

$\chi^2_\nu$ & 1.22  & 0.96 & 0.92 & 1.26  & 1.09 \\

dof & 92 & 90 & 92 & 97 & 75 \\
  \hline
  \hline
  \multicolumn{6}{c}{BB+GAUSS} \\
  \hline


   $kT_{\rm BB}$ (eV) & 83.0$^{+1.5}_{-0.9}$ & 82.4(9) & 82.2$^{+0.7}_{-0.6}$ & 82.7(8) & 82.4$^{+1.0}_{-0.9}$  \\

  $R_{\rm BB}$ (km) & 6.0(6) & 6.0$^{+0.6}_{-0.5}$ & 6.1$^{+0.6}_{-0.5}$ & 6.3$^{+0.6}_{-0.5}$ & 6.4(6)  \\

  $E_{\rm line}$ $^c$ (eV) & 745$^{+17}_{-27}$  & 745 & 745 & 745 & 745   \\

 $\sigma$ $^c$(eV) & 41.7$^{+51.3}_{-33.8}$ & 41.7 & 41.7 & 41.7 & 41.7   \\

 $|$Norm$|$ &  9.2$^{+3.5}_{-9.2}\times10^{-5}$  & $\leq$2.7$\times10^{-5}$ & $\leq$1.7$\times10^{-5}$ & $\leq$2.9$\times10^{-5}$ &$\leq$2.4$\times10^{-5}$  \\

 Eq. Width (eV) & 28$^{+9}_{-11}$ & 6$^{+9}_{-5}$ & $\leq$8 & $\leq$11 & $\leq$13 \\

  F-test ($10^{-3}$) & 3.5$\times 10^{-7}$ & 232 & 1000 & 750 & $>$1000 \\

  NHP $^d$ &  7.9$\times 10^{-1}$ & 6.0$\times 10^{-1}$ & 6.6$\times 10^{-1}$ & 3.6$\times 10^{-2}$ & 2.5$\times 10^{-1}$  \\

   $\chi^2_\nu$ & 0.88  & 0.95 & 0.93 & 1.27 & 1.11 \\

dof & 89 & 89 & 91 & 96 & 74\\
\hline \hline
\end{tabular}
\begin{list}{}{}

\item[$^{a}$] \nh\ was frozen at the value obtained for the phase averaged spectra: \nh\ = 1.9$\times10^{20}$\cm2 .
\item[$^{b}$] Fluxes are calculated in the 0.3--1.2 keV energy range and in units of (10$^{-11}$ \ergscm2 )
\item[$^{c}$] Line energy and width were frozen at the value obtained for the phase interval 0.1 -- 0.3: E$_{line}$ = 745 eV and $\sigma$ = 42 eV.
\item[$^{d}$]  NHP is the Null Hypothesis Probability.
\end{list}
\label{tab:all_pps}
\end{center}
\end{table*}

The same phase-averaged and phase-resolved analysis was then
repeated for all the remaining  \xmm\, observations in search for the presence of the 745\,eV absorption line. Since
observations in Full Frame mode were affected by pile-up, their
spectra were extracted from an annular region extending from
15\arcsec\ to 30\arcsec\ from the nominal source position, this
time including also events with PATTERN$\leq$4. The resulting
values of the line equivalent width and the F-test probability as
a function of time are shown in Fig. \ref{fig:ftest} (top
panel), and clearly indicate that the presence of the feature is
unconstrained in most of the observations, possibly because of
their lower statistics.

We then performed a simultaneous fit of the spectra in the
0.1--0.3 phase bin for all the observations in Full Frame mode
with the Thin filter (18 spectra in total; see Table\,1). We
constrained \nh\, and the line energy to be the same (to minimize
the free parameters), and the width of the feature was frozen at
the value obtained for the May 2003 observation. The
phase-dependent feature is detected at an energy of 785(13) eV for
the \texttt{tbabs*(bbodyrad+gauss)} model (\rchisq\ = 0.96 for 556 dof); normalizations
and equivalent widths are consistent within the errors in all the
spectra. The \texttt{tbabs*bbodyrad} model gives \rchisq\
= 1.03 for 575 dof. The F-test probability for the inclusion of
the Gaussian absorption line is 4.2$\times10^{-6}$, which
corresponds to a significance level of $\sim4.5\sigma$ (taking
into account the number of trials).

To further test the presence of this feature in data taken before
and after May 2nd, 2003, we also combined all the spectra except
that of May 2nd, 2003. We used the {\tt SAS epicspeccombine} tool
to merge the spectra and the response matrices. For the combined
spectrum in the phase range 0.1--0.3, the best-fit with a
blackbody model yields \rchisq\ = 2.04 for 123 dof; by including a
Gaussian absorption line, \rchisq\ decreases to 1.48 for 121 dof.
The parameters for the phase-dependent line are: E$_{line}$ =
795$^{+15}_{-14}$ eV, normalization of -8.3$^{+1.7}_{-1.7}$ and
equivalent width of 27$^{+6}_{-5}$\,eV for the \texttt{gauss}
model (the line width was fixed to the value obtained from the May 2nd, 2003 observation). The F-test probability between these
two models is 1.3$\times10^{-9}$ and the significance is equal to
6$\sigma$ (again considering 5 trials).

As a further check, we built a combined spectrum comprising
all observations. This spectrum, created by merging
all the spectra extracted from a 15\arcsec\ --
30\arcsec\, annular region with PATTERN$\leq$4, has an exposure time of
406.8\,ks and a much improved statistics. The 0.1-0.3
phase-resolved spectrum shows an
absorption feature at energy 787$^{+15}_{-14}$ eV with
normalization of -7.9$^{+1.6}_{-1.7} \times 10^{-5}$ and
equivalent width of 26$^{+4}_{-5}$\,eV for the \texttt{gauss}
model; also in this case we froze the line width at the value
estimated for the longest observation. The F-test probability for
the inclusion of the absorption feature is now 4.7$\times 10^{-9}$
and, taking into consideration the number of trials, the
significance level is again $\sim6\sigma$. As shown in Fig.
\ref{fig:fit}, panel d of, the spectral feature is clearly detectable
in the fit residuals of the merged observations, and the line
significance increases with respect to the May 2003 observation
alone.

In the spectra relative to the other phase intervals the line
equivalent width is always $<13$ eV, and the F-test probability is
never $<3\times10^{-3}$ (see Fig. \ref{fig:ftest}, lower panel).
We stress that the response of the pn camera has slightly changed (see, e.g., Sartore et al. 2012) over the twelve
years spanned by the observations, hence mixing spectra obtained
with different settings over such a long period certainly
introduce large systematic errors which are difficult to quantify.
However, a spectral feature present only in a given phase interval
cannot result from systematic effects, which are definitely
independent on the rotational phase. While the line is mostly
evident in the May 2nd, 2003 observation, which we used to derive
the best estimate of the line parameters, the considerations
presented above strongly suggest that the feature was present also
in the other observations, with properties which appear relatively
stable over the 12 yr period covered by \xmm\, data.

\section{Discussion}
\label{discussion}

A careful re-analysis of all available \xmm\ observations of the
isolated neutron star \xin\ revealed the presence of a
phase-dependent absorption feature at $\sim750$\,eV, which is
mostly evident during the decline of the pulse profile (phases
0.1--0.3). The feature was first significantly detected (at $\sim
5\sigma$ confidence level) in the May 2nd, 2003 \xmm\ observation,
which was the longest one and was less affected by
pile-up. Phase-dependent spectral analysis of the remaining
datasets, which cover the period May 2000--September 2012,
indicates that the feature is likely present at all epochs and its
properties are consistent with being constant over the timespan
covered by available observations ($\sim 12$ yrs). This suggests
that the feature may be long-lasting and not associated to the
timing anomalies (possibly due to a glitch) which occurred in 2003 (van
Kerkwijk et al. 2007; Hohle et al. 2012b).

Interestingly, an absorption feature with somewhat similar
properties has been discovered in the spectrum of the
``low-field'' magnetar \src\ ($B_{\rm dip}\sim 6\times10^{12}$\,G;
Tiengo et al. 2013). In both cases the feature is present only in
a quite narrow phase interval ($\sim 1/5$ of the phase cycle), but in
\src\ the line energy is higher ($\gtrsim 2$ keV) and strongly
variable in phase (by a factor $\sim 5$ in one tenth of the phase cycle). No clear variation of the line energy with phase is seen
in \xin, although this might simply reflect the fact that the
source counts become background dominated at energies just above
that of the feature. Tiengo et al. (2013) interpreted the feature
in \src\ as due to proton cyclotron resonant scattering in a
baryon-loaded magnetic loop close to the star surface. The
$B$-field variation along the loop explains the strong dependence
of the line energy with phase, since photons emitted from the
surface intersect the loop at different positions.

The similarities between the two features are suggestive that the
same mechanism may be at work in \xin. If the feature is a
(broadened) proton cyclotron line, its energy is  $E_{\rm line} =
0.63 (B_{\rm loop}/10^{14}\, {\rm G})/(1+z)$\, keV, where $1/(1+z)
\sim 0.7$ is the gravitational redshift at the star surface. The
implied magnetic field in the loop is $B_{\rm loop}\sim
1.8\times10^{14}$\,G, a factor of $\sim7$ larger than the the
equatorial value of the dipole  magnetic field at the surface,
$B_{\rm dip}\sim 2.5\times10^{13}$\,G.  A preliminary calculation
using the same geometrical model as in Tiengo et al. (2013) shows
that the energy of the line and its appearance only for $\sim
20$\% of the star period can indeed be reproduced. It should be
noted, however, that the exact geometry of the magnetic structure
is likely different in the two sources. In the case of \src, in
fact, X-ray photons come from a quite small, hot spot with size
$\sim 1$\, km, while the emitting region of \xin, albeit not
uniform, is larger ($R_{\rm BB}\sim 6$\, km).

A further possibility is that the line results from atomic
absorption in the neutron star (magnetized) atmosphere. The strong
dependence of the feature on the star rotational phase may be due
to local inhomogeneities in the star surface magnetic field (as in
the cyclotron scenario above) and/or to a clumpy structure of the
atmosphere itself. The possibility that primary photons come from
a small hot spot appears, on the other hand, unlikely, given the
relatively large radiation radius and the broad shape of the pulse
profile. If the feature arises from  bound-bound (or bound-free)
transitions, the atmosphere composition is most probably He or
mid-Z elements (see, e.g., Pavlov \& Bezchastnov 2005; Mori \& Ho
2007). The ionization energy for H is, in fact, below $\sim 0.7$
keV for fields $\la 10^{14}\ \mathrm G$ (above this value the
feature may be washed out by vacuum resonance mode conversion,
e.g. Van Kerkwijk \& Kaplan 2007 and references therein).

The presence of proton cyclotron lines produced by resonant
scattering/absorption in confined, high-$B$ structures close to
the star surface, where the magnetic field is a factor $\gtrsim
10$ higher than the dipole, would be supportive of a picture
in which the magnetic field of (highly magnetized) neutron stars
is complex, with substantial deviations from a pure dipole on the
small scales. In recent years, a better (although far from
conclusive) understanding of how stellar magnetic fields are
generated and amplified indicates that strong, non-dipolar field
components are ubiquitous, from massive stars (Braithwaite \&
Spruit 2005) to proto-neutron stars (Obergaulingiler, Janka \&
Aloy 2014). It is now well established that internal toroidal
field components, multipolar surface structures, as well as very
localized B-field bundles must be present in many neutron stars,
especially in highly-magnetized ones. The recent discovery of
sources showing strong magnetic-powered flares and outbursts, but
with low dipolar magnetic fields, $\lesssim 10^{13}$\,G,  further
strengthens this idea (Rea et al. 2010, 2012, 2014; Turolla \&
Esposito 2013).

Finally, we stress that the absorption feature reported here for
the first time is unlikely related to the broad absorption feature
at 270\,eV (Haberl et al. 2004, 2006). Even if the two may have the same physical
origin, i.e. a proton cyclotron resonance, the latter implies a
lower magnetic field, $\sim 5\times 10^{13}$\, G, comparable to
the spin-down estimate, $B_{\rm dip}\sim 2.4\times 10^{13}$\, G,
indicating that in this hypothesis it is related to the
large-scale field. Actually broad absorption features have been
detected in all the XDINSs, with the exception of RX J1856.5-3754
(e.g. Turolla 2009), but the strong phase variability we observe for the 0.75 keV line of \xin\ sets it apart.

This is the first time evidence of a complex magnetic field
structure is observed in an XDINS, although this result is not
unexpected, if, as suggested by many investigations, XDINSs are
aged magnetars (Vigan\`o et al. 2013; see also Turolla 2009).

\acknowledgments

AB, NR and FCZ are supported by an NWO Vidi Grant (PI: Rea) and by
the European COST Action MP1304 (NewCOMPSTAR). NR acknowledges
support by grants AYA2012--39303 and SGR2014--1073. The work of RT
and AT is partially supported by an INAF PRIN grant.





\end{document}